\documentclass{ws-procs975x65}
 \usepackage{wrapfig}
\def\beq{\begin{equation}}
\def\eeq{\end{equation}}

\begin{document}

\title{Motion of a spinning particle in curved space-time}
\author{S. Satish Kumar$^*$}

\address{Lorentz Institute, Leiden University\\
Niels Bohrweg 2, Leiden, Netherlands\\
$^*$E-mail: satish@lorentz.leidenuniv.nl\\}

\begin{abstract}
The motion of spinning test-masses in curved space-time is described with a covariant hamiltonian formalism.
A large class of hamiltonians can be used with the model-independent Poisson-Dirac brackets, to obtain  equations of motion. Here we apply it to  the minimal  hamiltonian and also to a non-minimal hamiltonian, describing the gravitational Stern-Gerlach force.  And a note on $ISCO$ has been added. 
\end{abstract}


\bodymatter

\section{Spinning particles}
The study of test-mass with intrinsic angular momentum or spin and its dynamics in curved space-time, was very essential since the beginning of General Relativity. There are two complimentary approaches to the subject. Since the gravitating objects possess quasi-rigid rotation along with orbital motion,  studies have aimed at keeping track of the centre of mass by using different supplementary conditions with in the Mathisson-Papapetrou model \cite{math1937, 1951pa, kh89, sem99, sch99,  kysem07, bara09, ply11, ste04, cos15, kh99, bini}.

In practise, determining the overall motion of the body, by following a detailed microscopic description of a material body is often too complicated. Therefore the spinning particle approximation, in contrast neglects the internal structure by assigning an overall position, momentum and spin to a test mass moving on a worldline.

\section{Covariant Hamiltonian Formalism}
In recent papers \cite{sat15a, sat15b, vholten15} we derived equations of motion for compact spinning bodies in curved space-time in an effective world-line formalism. The equations can be obtained either in a hamiltonian formulation 
or from local energy-momentum conservation. In the hamiltonian formulation, the dynamical systems are specified by the following three sets of ingredients:

 \subsection{Equations of motion}
In the simplest case of a massive free spinning particle in the absence of Stern-Gerlach forces and external fields the equations read
\begin{equation}
\begin{array}{l}
\displaystyle{D_{\tau} u^{\mu}  = \frac{1}{2m}\, \Sigma^{\kappa\lambda}\, R_{\kappa\lambda\;\,\nu}^{\;\;\;\,\mu} u^{\nu}}, \qquad 
\displaystyle{D_{\tau} \Sigma^{\mu\nu} =  0. } 
\end{array}
\label{2.1a}
\end{equation}
The solution of these equations  is the worldline (as shown in Fig.1) along which the spin tensor is covariantly constant rather than a worldline followed by some preferred centre of mass, however chosen. \\
\begin{wrapfigure}{r}{0.16\textwidth}
  \begin{center}
   \includegraphics[width=0.16\textwidth]{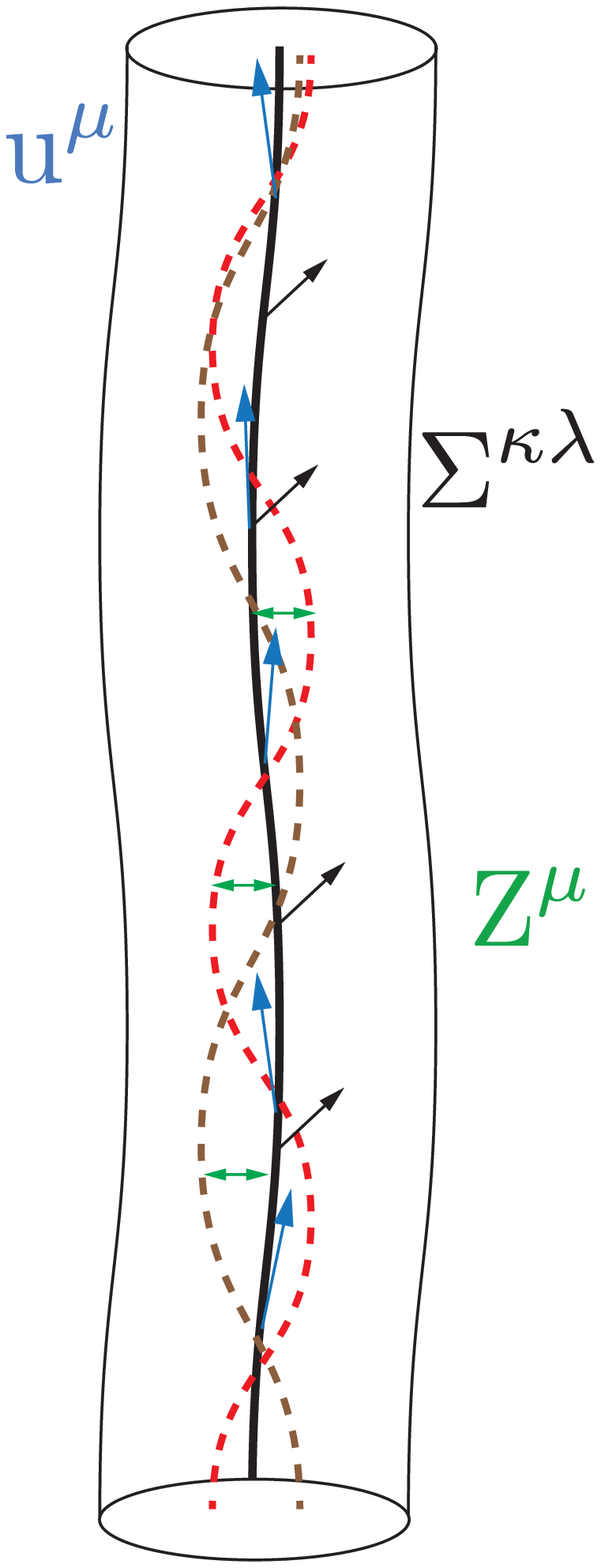}
   \vspace{-25pt}
  \end{center}
        \caption{}
\end{wrapfigure}%
Here $m$ is the mass of the spinning particle, $\tau$ is the proper time, and  the proper velocity $u^{\mu} = \dot{x}^{\mu}$ is the tangent vector (blue arrows) of this worldline (thick, black line), which may differ from the proper velocity of any centre of mass (dotted lines) and the difference is quantified by the dipole vector $Z^{\mu}$ (green, double arrows):
\begin{equation}
S_{\mu} = \frac{1}{2}\, \sqrt{-g}\, \varepsilon_{\mu\nu\kappa\lambda} u^{\nu} \Sigma^{\kappa\lambda}, \qquad
Z_{\mu} = \Sigma_{\mu\nu} u^{\nu}.
\label{2.1b}
\end{equation}
where $\Sigma^{\mu\nu}$ is the anti-symmetric spin tensor which unifies the internal angular 
momentum pseudo-vector $S^{\mu}$ and the mass dipole moment $Z^{\mu}$.
Both vectors are space-like:
\begin{equation}
S_{\mu} u^{\mu} = Z_{\mu} u^{\mu} = 0,
\label{2.1c}
\end{equation}
as required to match the number of independent components of $\Sigma^{\mu\nu}$. 

\subsection{One-particle phase space}
In the phase-space formulation the equations of motion (\ref{2.1a}) can be derived from the free hamiltonian
\begin{equation}
H_0 = \frac{1}{2m}\, g^{\mu\nu} \pi_{\mu}\pi_{\nu}, \qquad \pi_{\mu} = m g_{\mu\nu} u^{\nu}, 
\label{2.2a}
\end{equation}
by applying the model-independent Dirac-Poisson brackets
\begin{equation}
\begin{array}{l}
\displaystyle{ \left\{ x^{\mu}, \pi_{\nu} \right\} = \delta_{\nu}^{\mu}, \qquad 
\left\{ \pi_{\mu}, \pi_{\nu} \right\} = \frac{1}{2}\, \Sigma^{\kappa\lambda} R_{\kappa\lambda\mu\nu}, }\\
 \\
\displaystyle{ \left\{ \Sigma^{\mu\nu}, \pi_{\lambda} \right\} = \Gamma_{\lambda\kappa}^{\;\;\;\mu}\, \Sigma^{\nu\kappa}  
 - \Gamma_{\lambda\kappa}^{\;\;\;\nu}\, \Sigma^{\mu\kappa}, }\\
 \\
\displaystyle{ \left\{ \Sigma^{\mu\nu}, \Sigma^{\kappa\lambda} \right\} =  g^{\mu\kappa} \Sigma^{\nu\lambda} - g^{\mu\lambda} \Sigma^{\nu\kappa}
 - g^{\nu\kappa} \Sigma^{\mu\lambda} + g^{\nu\lambda} \Sigma^{\mu\kappa}. }
\end{array}
\label{2.2b}
\end{equation}
Here the inverse metric $g^{\mu\nu}$, the connection $\Gamma_{\mu\nu}^{\,\,\,\,\,\lambda}$ and the Riemann tensor $R_{\mu\nu\kappa\lambda}$ appear as structure functions in the bracket algebra. 

\subsection{Constants of motion}
This formalism is also convenient for deriving constants of motion, which commute with the hamiltonian in the sense of the brackets. For a test mass moving on spherically symmetric space-times, there exist conservation laws as implied by Noether's theorem. The symmetries of the space-time manifest themselves as Killing vectors. 

Here due to spin-orbit coupling, the conserved quantities implied by Noether's theorem are linear combinations of momentum and spin components:
\begin{equation}
\{J, H_{0}\} = 0 \quad \Rightarrow \quad J = {\alpha}^{\mu}\, \pi_{\mu} + \frac{1}{2}\,\beta_{\mu\nu}\, \Sigma^{\mu\nu}.
\label{2.3a}
\end{equation}
Such a constant exists if 
\begin{equation}
\nabla_{\mu}\alpha_{\nu} + \nabla_{\nu}\alpha_{\mu} = 0, \qquad \nabla_{\lambda}\beta_{\mu\nu} = R_{\mu\nu\lambda}^{\,\,\,\,\,\,\,\,\,\kappa} \alpha_{\kappa}.
\label{2.3b}
\end{equation}
Thus $\alpha$ is a Killing vector and $\beta$ is its anti-symmetrized gradient:
\begin{equation}
\beta_{\mu\nu} = \frac{1}{2}\, (\nabla_{\mu}\alpha_{\nu} - \nabla_{\nu}\alpha_{\mu} ).
\label{2.3c}
\end{equation}
In addition, for any space-time geometry there exist universal constants of motion:  the particle's mass given by the hamiltonian constraint; the total spin $I$ as a result of local Lorentz invariance; and the pseudo-scalar spin-dipole product is defined as
\begin{equation}
H_{0} = -\frac{m}{2},  \qquad I = \frac{1}{2}\,g_{\mu\kappa}g_{\nu\lambda}\Sigma^{\mu\nu}\Sigma^{\kappa\lambda},  \qquad  D= \frac{1}{8} \sqrt{-g} \,\epsilon_{\mu\nu\kappa\lambda}\Sigma^{\mu\nu}\Sigma^{\kappa\lambda}.
\label{2.3d}
\end{equation}

\section{Schwarzschild space-time}
The Schwarzschild metric is the unique, static and spherically symmetric vacuum solution, as implied by Birkhoff's theorem. Its standard line element in Droste co-ordinates reads:
\begin{equation}
d\tau^2 = \left( 1 - \frac{2M}{r} \right) dt^2 - \frac{dr^2}{1 - \frac{2M}{r}} - r^2 d\theta^2 - r^2 \sin^2 \theta\, d\varphi^2.
\label{4a}
\end{equation}
As a result of the the time-translation symmetry there is a Killing vector field corresponding to the particle energy
\begin{equation}
E = - \pi_t - \frac{M}{r^2}\, \Sigma^{tr}.
\label{4b}
\end{equation}
The spherical symmetry implies three Killing vector fields generating a conserved angular momentum 
3-vector 
\begin{equation}
\begin{array}{lll}
J_1 & = & - \sin{\varphi}\,\pi_{\theta} - \cot{\theta} \cos{\varphi}\, \pi_{\varphi} \\
 & & \\
 & &  - r \sin \varphi\, \Sigma^{r\theta} - r \sin \theta \cos \theta \cos \varphi\, \Sigma^{r\varphi} + r^2 \sin^2 \theta \cos \varphi\, \Sigma^{\theta\varphi}, \\
 & & \\
J_2 & = &  \cos \varphi\, \pi_{\theta} - \cot \theta \sin \varphi\, \pi_{\varphi} \\
 & & \\
 & &  + r \cos \varphi\, \Sigma^{r\theta} - r \sin \theta \cos \theta \sin \varphi\, \Sigma^{r\varphi} + r^2 \sin^2 \theta \sin \varphi\, \Sigma^{\theta\varphi}, \\
 & & \\
J_3 & = &  \pi_{\varphi} + r \sin^2 \theta\, \Sigma^{r\varphi} + r^2 \sin \theta \cos \theta\, \Sigma^{\theta\varphi}. 
\end{array}
 \label{4c}
\end{equation}
The invariance under rotations allows us to choose the direction of total angular momentum to be the $z$-direction, 
such that
\begin{equation}
{\bf{J}} = (0, 0, J).
\label{4d}
\end{equation}
Therefore we can express four spin-tensor components in terms of co-ordinates and velocities: 
\begin{equation}
\begin{array}{ll}
\displaystyle{\Sigma^{r\theta}} = \displaystyle{- mr u^{\theta}}, & \qquad \displaystyle{\Sigma^{r\varphi}} = \displaystyle{\frac{J}{r} - m r u^{\varphi}, } \\
 & \\
\displaystyle{\Sigma^{\theta\varphi}} = \displaystyle{\frac{J}{r^2}\, \cot \theta,} & 
\qquad \displaystyle{\Sigma^{tr}} = \displaystyle{\frac{mr^2}{M} \left( 1 - \frac{2M}{r} \right) u^t - \frac{r^2 E}{M}. } 
\end{array}
 \label{4e}
\end{equation}
The two remaining spin-tensor components $\Sigma^{t\theta}$ and $\Sigma^{t\varphi}$ are related (\ref{2.3d}) to the constants of motion $I$ and $D$ and the orbital parameters. In addition, we have the hamiltonian constraint: $u^{2} = - 1:$
\begin{equation}
 \left( 1 - \frac{2M}{r} \right) u^{t\,2} - \frac{u^{r\,2}}{1 - \frac{2M}{r}}  - r^2 u^{\theta\,2}- r^2 \sin^2 \theta\, u^{\varphi\,2} = 1.
 \label{4f}
\end{equation}
The total angular momentum $J$ of spinless particles is strictly orbital, therefore they move in a plane perpendicular to the angular momentum. With our choice of the $z$-axis this is the equatorial place $\theta = \pi/2$. For the spinning particles, as the precession of the spin is compensated by the precession of the orbital angular momentum and they are not necessarily aligned, the plane of the orbit can precess \cite{inprep,bini}.

\section{Plane-circular orbits}
To find the simplest orbit: circular, we consider the motion to be in the equatorial plane. Then the planar motion \cite{sat15a} requires alignment of the spin with the orbital angular momentum  in the Schwarzschild geometry. Which implies, \\
\begin{equation}
\theta = \frac{\pi}{2} \quad \Rightarrow  \quad u^{\theta} =  \dot{u}^{\theta} = 0,
 \label{5a}
\end{equation}
as a result
\begin{equation}
\Sigma^{t\theta} = \Sigma^{r\theta} = \Sigma^{\theta\varphi} = 0.
 \label{5b}
\end{equation}
Such orbits satisfy the conservation law 
\begin{equation}
D = S  \cdot  Z = 0.
 \label{5c}
\end{equation}
For circular orbits we additionally impose $r = R = \text{constant}$ and $u^{r} =  \dot{u}^{r} = 0.$ Then the equations of motion (\ref{2.1a}) implies $(\Sigma^{tr}, \Sigma^{r\varphi}, \Sigma^{t\varphi})$ are constants. Further, the symmetry of the circular orbit guarantees that $(u^{t}, u^{\varphi})$ are constant in time, and therefore $\Sigma^{t\varphi} = 0$ indeed. 

Since the radial acceleration is absent $D_\tau u^{r} = 0$ and using the hamiltonian constraint (\ref{4f}), the equation of motion which describes the circular orbit in terms of constants $(R, J)$ is found to be
\begin{equation}
\frac{J M}{m R^2} \left( \frac{2M}{R} + R^2 u^{\varphi\,2}  \right) = R u^{\varphi} \left[ \frac{M}{R} - \left( 1 - \frac{6M}{R} + \frac{6M^2}{R^2} \right) R^2 u^{\varphi\,2}   \right],
 \label{5d}
\end{equation}
which can be solved for orbital frequency $u^{\varphi}$ (therefore $u^{t}$).  It can  also be  written equivalently in terms of other constants $(R, E)$, by using the relation between $J$ and $E$ obtained by setting the condition: the rate of change of $\Sigma^{t\varphi}$ vanishes, as implied by circular orbits. 
Finally the total spin for  circular orbits written in terms of constants $(R, J, E)$:
\begin{equation}
I = - \frac{R^4}{M^2} \left[ \left( 1 - \frac{2M}{R} \right) m u^{t} - E  \right]^2 + \frac{1}{\left( 1 - \frac{2M}{R} \right)} \left[ J - m R^2 u^{\varphi} \right]^2.
 \label{5e}
\end{equation}

\newpage
\subsection{Innermost Stable Circular Orbit (ISCO)}
\begin{wrapfigure}{r}{0.39\textwidth}
\vspace{-20pt}
  \begin{center}
   \includegraphics[width=0.39\textwidth]{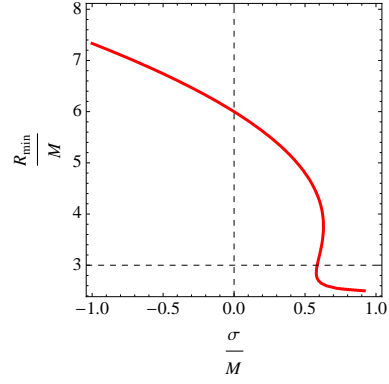}
    \vspace{-20pt}
  \end{center}
        \caption{The radius of minimal orbital angular momentum $R_{min}/M$ at fixed spin.}
\end{wrapfigure}
Re-writing (\ref{5d}) with orbital angular momentum $\ell = R^2 u^{\varphi}$ and spin parameter $\sigma = R\,\Sigma^{r\varphi} / m$, in the form 
\begin{equation}
\hspace{-40mm}\frac{\sigma}{M}\frac{R}{M} \left( \frac{2R}{M} + \frac{\ell^2}{M^2} \right) \nonumber
\end{equation}
\begin{equation}
\hspace{-5mm}= \frac{\ell}{M} \left(\frac{R}{M} - 2 \right) \left[ \frac{R^2}{M^2} -  \frac{\ell^2}{M^2} \left(\frac{R}{M} - 3\right) \right],
 \label{4.1}
\end{equation}
we derive the minimum of $\ell/M$ as a function of $R/M$ for fixed spin $\sigma$ and the result if plotted in Fig. 2. It coincides with the $ISCO$ in the absence of spin: $R=6M$ when $\ell^2 = 12 M^2$. We have established that the curve approximates the $ISCO$ well in the domain of physical interest $-0.5\, < \, \sigma/M \, < \, 0.5$.  Beyond this limit, the back reaction of the spinning particle on the space-time geometry can no longer be neglected anyway. We have also presented the detailed analysis of $ISCO$ from the arguments of stability criterion  \cite{sat15b}.

\section{Gravitational Stern-Gerlach force}
We can extend our analysis by including a spin-spin interaction, with coupling constant $\kappa$:
\begin{equation}
H = H_{0} + H_{SG}, \qquad H_{SG} = \frac{\kappa}{4}\,R_{\mu\nu\kappa\lambda} \Sigma^{\mu\nu} \Sigma^{\kappa\lambda},
\end{equation}
Using this non-minimal hamiltonian in the brackets (\ref{2.2b}), we generalize the equations of motion (\ref{2.1a}) and obtain the gravitational Stern-Gerlach force, coupling spin to the gradient of the curvature
\begin{equation}
\begin{array}{l}
\displaystyle{\pi_{\mu} = m g_{\mu\nu} \dot{x}^{\nu}, \qquad D_{\tau} \pi_{\mu}  = \frac{1}{2m}\, \Sigma^{\kappa\lambda}\, R_{\kappa\lambda\mu}^{\;\;\;\;\;\,\,\nu} \pi_{\nu} - \frac{\kappa}{4}\,\Sigma^{\kappa\lambda} \Sigma^{\rho\sigma} \nabla_{\mu} R_{\kappa\lambda\rho\sigma}} , \\
 \\
\qquad \qquad \qquad \qquad \displaystyle{D_{\tau} \Sigma^{\mu\nu} =  \kappa \Sigma^{\kappa\lambda} \left( R_{\kappa\lambda\;\;\sigma}^{\;\;\;\,\,\mu}\Sigma^{\nu\sigma} - R_{\kappa\lambda\;\;\sigma}^{\;\;\;\,\,\nu}\Sigma^{\mu\sigma} \right) }.
\end{array}
\end{equation}
Remarkably all the conservation laws established for the minimal hamiltonian holds true here as well.

\section{Summary} 

A covariant hamiltonian spin-dynamics in curved space-time is formulated without using any supplementary conditions like Pirani, Tulczyjew and so on. As a result the Pirani vector $Z$ is no longer zero in the theory. The advantage of this formalism is, that the closed set of model-independent Poisson-Dirac brackets can be used with different hamiltonians, using the minimal choice of hamiltonian we obtain the equations of motion by computing its bracket with this hamiltonian. The particle's motion along Killing trajectories, like circular orbits have been studied. Next the study has been extended with gravitational Stern-Gerlach interactions by introducing the non-minimal hamiltonian.

\section{Acknowledgements} I am grateful to my collaborators of this work, G. d'Ambrosi and J. W. van Holten. I thank the organisers of the Fourteenth Marcel Grossmann Meeting - MG14, for giving me an opportunity to present this work. I also acknowledge NIKHEF for all the hospitality extended to me during this project.


\begin{thebibliography}{10}


\bibitem{math1937} 
M.~Mathison, {\em Acta Phys. Polon.}  {\bf{6}} (1937) 163

\bibitem{1951pa} 
A.~Papapetrou, {\em Proc. Roy. Soc. Lond.}  {\bf{A209}} (1951) 248-258

\bibitem{kh89} 
I.~Khriplovich, {\em Sov. Phys.} JETP   {\bf{69}} (1989), 217

\bibitem{sem99} 
O.~Semerak, {\em Mon. Not. Roy. Astron. Soc. }  {\bf{308}} (1999), 863


\bibitem{sch99} 
G.~Schaefer, {\em Gen. Rel. Grav.}  {\bf{36}} (2004), 2223

\bibitem{kysem07} 
K.~Kyrian, and O.~Semerak, {\em Mon. Not. Roy. Astron. Soc.} {\bf {382}} (2007), 1922


\bibitem{bara09} 
E.~ Barausse, E.~Racine, and A.~Buonanno, {\em Phys. Rev. D} {\bf {80}} (2009), 104025; {\tt arXiv:0907.4745 [gr-qc]}.

\bibitem{ply11} 
R.~ Plyatsko, O.~Stephanyshyn, and M.~Fenyk, {\em Class. Quan. Grav. } {\bf {28}} (2011), 195025; {\tt arXiv:1110.1967 [gr-qc]}.

\bibitem{ste04} 
J.~Steinhoff, {\em Ann. d. Physik}  {\bf{523}} (2011), 296; {\tt arXiv:1106.4203 [gr-qc]}.

\bibitem{cos15} 
L.~F.~ Costa, and J.~Natario,  {\em Fundamental Theories of Physics } {\bf {179}}  (2015), 215; {\tt arXiv:1410.6443 [gr-qc]}.


\bibitem{kh99} 
I.~ Khriplovich, and A.~Pomeransky,  {\em Surveys High En. Phys.} {\bf {14}} (1999), 145; {\tt arXiv:gr-qc/9809069}.


\bibitem{bini} 
D.~Bini, A.~Geralico, and R.~T.~Jantzen, {\em Class. Quant. Grav.,} {\bf {43}} (2011) 959, {\tt arXiv:1408.4946 [gr-qc]}.

\bibitem{sat15a} 
G.~d'Ambrosi, S.~Satish Kumar and J.~W.~van Holten, {\em Phys. Lett.} {\bf{B743}} (2015) 478,   {\tt arXiv:1501.04879v2 [gr-qc]}.

\bibitem{sat15b} 
G.~d'Ambrosi, S.~Satish Kumar,  J.~ van de Vis, and J.~W.~van Holten,  {\tt arXiv:1511.05454v1 [gr-qc]}.

\bibitem{vholten15} 
J.~W.~van Holten,  {\tt arXiv:1504.04290v2 [gr-qc]}.

\bibitem{inprep} 
S.~Satish Kumar and J.~W.~van Holten, {\em in preparation} (2015). 

\end{thebibliography}
\end{document}